\documentclass[aps,prb,twocolumn,showpacs,groupedaddress]{revtex4-1}
\usepackage{graphicx}

\begin{document}

\title{Robust Ferroelectric State in Multiferroic Mn$_{1-x}$Zn$_x$WO$_4$}
\author{R. P. Chaudhury$^{1}$, F. Ye$^2$, J. A. Fernandez-Baca$^{2,3}$, B. Lorenz$^{1}$, Y. Q. Wang$^{1}$, Y. Y. Sun$^{1}$, H. A. Mook$^2$, and C. W. Chu$^{1,4}$}
\affiliation{$^{1}$TCSUH and Department of Physics, University of Houston, Houston, TX 77204-5002}
\affiliation{$^2$ Neutron Scattering Science Division, Oak Ridge National Laboratory, Oak Ridge, TN 37831-6393, USA}
\affiliation{$^3$ Department of Physics and Astronomy, The University of Tennessee, Knoxville, TN 37996-1200, USA}
\affiliation{$^{4}$Lawrence Berkeley National Laboratory, 1 Cyclotron Road, Berkeley, CA 94720}
\date{\today }

\begin{abstract}
We report the remarkably robust ferroelectric state in the multiferroic compound Mn$_{1-x}$Zn$_x$WO$_4$. The substitution of the magnetic Mn$^{2+}$ with nonmagnetic Zn$^{2+}$ reduces the magnetic exchange and provides control of the various magnetic and multiferroic states of MnWO$_4$. Only 5 \% of Zn substitution results in a complete suppression of the frustrated collinear (paraelectric) low temperature phase. The helical magnetic and ferroelectric phase develops as the ground state. The multiferroic state is stable up to a high level of substitution of more than 50 \%. The magnetic, thermodynamic, and dielectric properties as well as the ferroelectric polarization of single crystals of Mn$_{1-x}$Zn$_x$WO$_4$ are studied for different substitutions up to x=0.5. The magnetic phases have been identified in single crystal neutron scattering experiments. The ferroelectric polarization scales with the neutron intensity of the incommensurate peak of the helical phase.
\end{abstract}

\pacs{75.30.-m,75.30.Kz,75.50.Ee,77.80.-e,77.84.Bw} \maketitle


%\preprint{}




%\thanks{}


%\keywords{}



\section{Introduction}
Multiferroic materials have received increasing attention in recent years because of the coexistence and mutual interaction of magnetic and ferroelectric orders and their coupling to external magnetic and electric fields.\cite{fiebig:05,spaldin:05,tokura:07} Improper ferroelectricity can be induced by certain kinds of inversion symmetry breaking magnetic order if the magnetic moments are strongly coupled to the lattice causing the ionic displacements with a macroscopic electrical polarization. The transverse helical magnetic structure was shown to be compatible by symmetry with a ferroelectric polarization arising from a third order coupling term of the ferroelectric and magnetic order parameters in the Ginzburg-Landau thermodynamic potential.\cite{kenzelmann:05,mostovoy:06} Helical (non-collinear) magnetic structures have in fact been found in various multiferroics, for example, in TbMnO$_3$,\cite{kenzelmann:05} Ni$_3$V$_2$O$_8$,\cite{lawes:05} and MnWO$_4$.\cite{taniguchi:06,heyer:06} Different microscopic models have been recently proposed to describe the phase complexity and ferroelectricity in multiferroic manganites.\cite{katsura:05,sergienko:06,sergienko:06b,mochiguzi:10}

The non-collinear magnetic order observed in these compounds is a consequence of magnetic frustration due to geometric constraints or competing exchange interactions resulting in a close competition of different magnetic structures that are nearly equal in energy. Most multiferroic materials are therefore extremely sensitive to small perturbations. Magnetic fields may stabilize or destroy the multiferroic state or result in the rotation of the ferroelectric polarization by 90$^{\circ}$ or even by 180$^{\circ}$.\cite{taniguchi:06,higashiyama:04,hur:04,seki:08} Similarly, the application of external pressure\cite{delacruz:07,delacruz:08,chaudhury:07,chaudhury:08b} as well as chemical substitutions\cite{chaudhury:08,chaudhury:09b,seki:07,kanetsuki:07} are a viable tool to change and tune the multiferroic properties of different compounds.

MnWO$_4$ is also known as the mineral H\"{u}bnerite. The crystallographic structure of the compound is monoclinic (space group: P 2/c). Three magnetic phase transitions at T$_N$=13.5 K, T$_C$=12.6 K, and T$_L$=7.8 K separate different collinear and non-collinear magnetic phases.\cite{lautenschlager:93} The AF3 phase (T$_C<$T$<$T$_N$) shows a sinusoidal spin order given by the incommensurate (IC) vector $\overrightarrow{q}_{3}=(-0.214,1/2,0.457)$. The collinear Mn spins are confined to the a-c plane forming an angle of about 34$^\circ$ with the a-axis. Below T$_C$ the spins tilt out of plane and form a helical, non-collinear structure breaking the spatial inversion symmetry (AF2 phase). The modulation vector of the spin order does not change at T$_C$, $\overrightarrow{q}_{2}$=$\overrightarrow{q}_{3}$. The helical phase becomes unstable at T$_L$ and the low temperature AF1 phase (ground state) is characterized by a frustrated $\uparrow\uparrow\downarrow\downarrow$ spin structure with a the commensurate (CM) modulation vector $\overrightarrow{q}_{1}=(-1/4,1/2,1/2)$. Among the three magnetic phases only the helical AF2 phase is ferroelectric. The details of the magnetic structures have been revealed in neutron scattering experiments.\cite{lautenschlager:93,sagayama:08} The complex phase diagram of MnWO$_4$ and the multiferroic properties have been described extensively through magnetic, dielectric, and thermodynamic measurements.\cite{ehrenberg:97,taniguchi:06,arkenbout:06,heyer:06,taniguchi:08,chaudhury:08c} Novel physical phenomena such as the coupling of ferroelectric with magnetic domains\cite{meier:09} or the control of the magnetic domain chirality by electric fields\cite{finger:10} have been reported.

The tuning of the magnetic exchange interactions and, consequently, the multiferroic properties by partial substitution of manganese by other transition metals, for example iron or cobalt, was achieved recently.\cite{obermayer:73,klein:74,garciamatres:03,ding:00,song:09} The magnetic phases and the multiferroic properties are extremely sensitive to small amounts of Fe substitution.\cite{chaudhury:08,ye:08,chaudhury:09b} Some of the observed effects of transition metal substitution can be qualitatively described by simple Heisenberg models with competing interactions and uniaxial anisotropy,\cite{lawes:04,chaudhury:08} however, a more quantitative and microscopic understanding of the complex magnetic interactions and the effects of Fe or Co substitution seems to be a challenge because of the fact that Fe as well as Co carry their own magnetic moments and their exchange interactions with the Mn spins as well as their anisotropy parameters are not known. To overcome some of these problems and to simplify the physics of magnetic exchange in substituted MnWO$_4$ we decided to replace the Mn$^{2+}$ by non-magnetic d-metals such as Zn$^{2+}$. We have therefore synthesized single crystals of Mn$_{1-x}$Zn$_x$WO$_4$ with x between 0 and 0.5 and investigated their magnetic, thermodynamic, and multiferroic properties.

\begin{figure}
\begin{center}
\includegraphics[angle=0,width=3in]{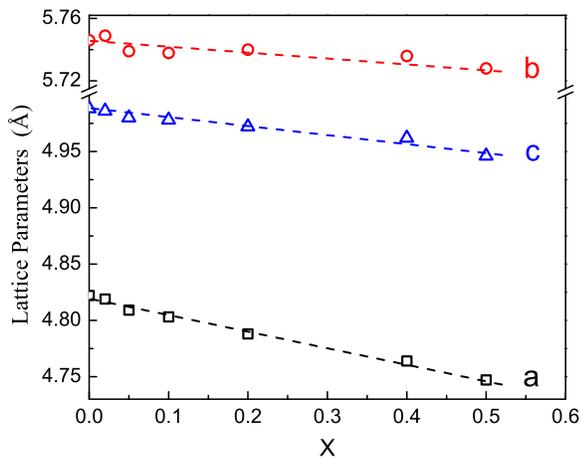}
\end{center}
\caption{(Color online) Lattice constants a, b, and c of Mn$_{1-x}$Zn$_x$WO$_4$ as function of x.}
\end{figure}

\section{Experimental}
The crystals were grown from a polycrystalline feed rod employing the floating zone optical furnace. The X-ray characterization shows that the a- and c-axes significantly decrease with the Zn substitution (Fig. 1) whereas the b-axis shows only a minor decrease within the resolution of the measurements. The smooth change of a, b, and c indicates that a solid solution of MnWO$_4$ and ZnWO$_4$ has formed. Small pieces were cut from the crystals and oriented using Laue single crystal X-ray techniques. The size and shape of the crystals was carefully chosen to fit the demands of the different experiments. The susceptibility was measured in a 5 Tesla Magnetic Property Measurement System (MPMS, Quantum Design). The ferroelectric polarization of the multiferroic phase was measured by integrating the pyroelectric current upon heating the sample in zero electric field at a speed of 1 K/min. Before the pyroelectric measurement the sample was cooled in an electric bias field of 3 kV/cm to align the ferroelectric domains. The heat capacity was measured in the Physical Property Measurement System (PPMS, Quantum Design). The elastic neutron scattering measurements were performed at the HB1A and HB1 three axis spectrometers at the High Flux Isotope Reactor at the Oak Ridge National Laboratory. The crystals were aligned in the scattering plane defined by the two orthogonal wave vectors (1,0,-2) and (0,1,0), in which the magnetic Bragg peaks and other structural peaks can be surveyed. The incident neutron energy was fixed at 14.7 meV using pyrolytic graphite crystals as monochromator, analyzer, and filter.

\begin{figure}
\begin{center}
\includegraphics[angle=0,width=3in]{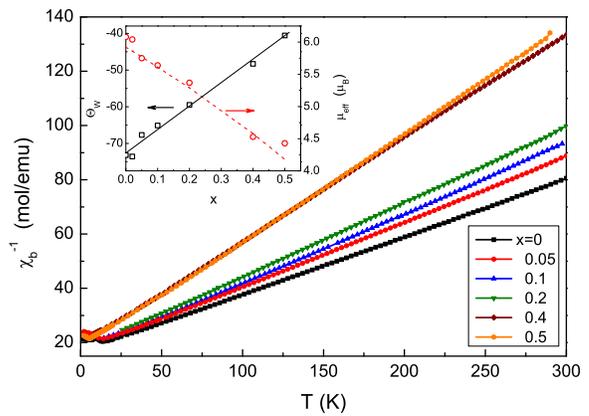}
\end{center}
\caption{(Color online) Inverse magnetic susceptibility of Mn$_{1-x}$Zn$_x$WO$_4$. The inset shows the x-dependence of the estimated effective magnetic moment $\mu_{eff}$ and the Curie-Weiss temperature $\theta_W$. The dashed line is the expected $\mu_{eff}$ for an Mn spin of S=5/2. The black (solid) line is a linear fit to the data of $\Theta_W$.}
\end{figure}

\section{Results and Discussion}
The b-axis magnetic susceptibility at high temperatures (T$>$50 K) follows the Curie-Weiss law, as shown in Fig. 2. From the inverse susceptibility the effective magnetic moment $\mu_{eff}$ per formula unit (Mn$_{1-x}$Zn$_x$WO$_4$) as well as the Curie-Weiss temperature $\Theta_W$ can be extracted and are displayed in the inset of Fig. 2. $\Theta_W$ is negative as expected for dominantly antiferromagnetic coupling and it's magnitude decreases with Zn substitution indicating the weakening of the magnetic coupling between the Mn spins. The effective magnetic moment also decreases with x since the Zn substitution continuously dilutes the system of Mn spins. The expected values for $\mu_{eff}$, assuming localized spins of the Mn$^{2+}$ ions with S=5/2, are shown by the dashed line in the inset of Fig. 2. The experimentally determined values for $\mu_{eff}$ are in very good agreement with the calculated values confirming the conclusion that the Zn$^{2+}$ substitutes for the Mn$^{2+}$ at levels up to the maximum of this work, 50 \%. This is further confirmed by Inductively Coupled Plasma Mass Spectrometry, testing the elemental composition at ten different spots of a single crystal with nominally 50 \% Zn doping. The average Zn concentration was determined as 0.49 $\pm 0.03$, close to the nominal composition.

\begin{figure}
\begin{center}
\includegraphics[angle=0,width=3in]{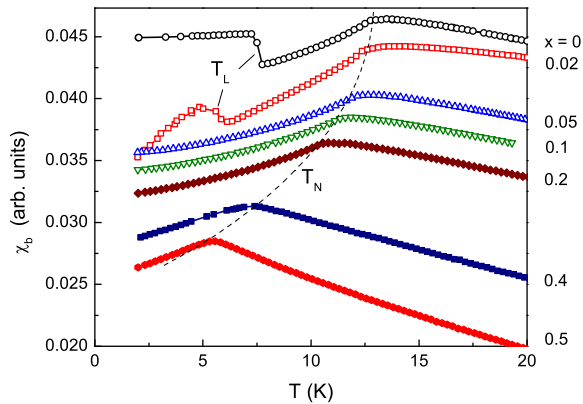}
\end{center}
\caption{(Color online) Low-temperature magnetic susceptibility $\chi_b$ of Mn$_{1-x}$Zn$_x$WO$_4$. Different curves are vertically offset for enhanced clarity.}
\end{figure}

The b-axis magnetic susceptibility at low temperatures clearly reveals the magnetic phase transitions at T$_N$ and T$_L$, as shown in Fig. 3 (different curves are vertically offset for better clarity). The transition into the AF3 phase is defined by the maximum of $\chi_b$(T) as indicated by the dashed line labeled T$_N$. With the lock-in transition into the AF1 phase $\chi_b$ shows a sharp increase at T$_L$ which is quickly shifted to lower temperatures with increasing x and is completely missing for x$\geq$0.05. 5\% of Zn substitution is sufficient to suppress the collinear AF1 phase and to stabilize the multiferrroic ferroelectric phase as the ground state. The effect of Zn substitution is opposite to the results of Fe substitution\cite{chaudhury:09b} where the multiferroic AF2 phase was suppressed and the low-T AF1 phase was stabilized with increasing Fe content. It is also interesting to note that a similar stabilization of the AF2 phase was reported for Co substituted MnWO$_4$ powder samples\cite{song:09} although the magnetic phases of Mn$_{1-x}$Co$_x$WO$_4$ are far more complex and several coexisting commensurate and incommensurate phase have been observed in single crystals at higher Co doping.\cite{chaudhury:10}

The main feature of the multiferroic phase is the existence of a spontaneous ferroelectric polarization that is oriented along the b-axis. Fig. 4 shows P$_b$(T) of Mn$_{1-x}$Zn$_x$WO$_4$ for x between 0 and 0.5 as measured by the pyroelectric current method. In perfect agreement with the magnetic data, the ferroelectric phase extends to the lowest temperature for x values equal or above 5 \%. The multiferroic phase becomes the ground state and it exists even at substitution levels as high as 50 \% while the onset temperature, T$_C$, is reduced with increasing x. It is interesting that the 2 \% substitution does decrease T$_L$ rapidly but there remains a finite polarization in the low-temperature state indicating the coexistence of the AF1 and AF2 phases. Similar phase coexistence has also been observed in recent single crystal neutron scattering experiments on Mn$_{1-x}$Fe$_x$WO$_4$\cite{ye:08,chaudhury:09c} and Mn$_{0.85}$Co$_{0.15}$WO$_4$.\cite{chaudhury:10} At higher substitution levels the ferroelectric polarization increases continuously with decreasing temperature. The polarization data of Fig. 4 prove that, unlike the Fe-substituted system Mn$_{1-x}$Fe$_x$WO$_4$, the substitution of the non-magnetic Zn ion favors the multiferroic ferroelectric state which becomes the ground state for x$\geq$0.05.

To identify the magnetic orders and their development with increasing Zn substitution neutron scattering experiments have been conducted. The temperature dependence of the integrated magnetic peak intensities in doped Mn$_{1-x}$Zn$_x$WO$_4$ is displayed in Fig. 5. The measurements were performed near the characteristic magnetic wavevector $\overrightarrow{q}_{2}=(-0.214,1/2,0.457)$ for the AF2 phase that is associated with the ferroelectric behavior. At the lowest Zn concentration x=0.02, the peak intensity first increases upon cooling below 13~K, and is significantly suppressed below 6~K but remains finite at lower temperature. Such behavior is accompanied by the sudden appearance of the commensurate magnetic order with $\overrightarrow{q}_{1}=(-0.25,0.5,0.5)$ below T$_{L}$=6~K (solid circles in Fig.~5). For higher Zn-doping systems, only incommensurate magnetic orders exist at all temperatures, and the thermal evolution of the integrated intensity closely tracks the polarization measurement evidencing the intimate connection between the spiral order and the ferroelectricity.

\begin{figure}
\begin{center}
\includegraphics[angle=0,width=3in]{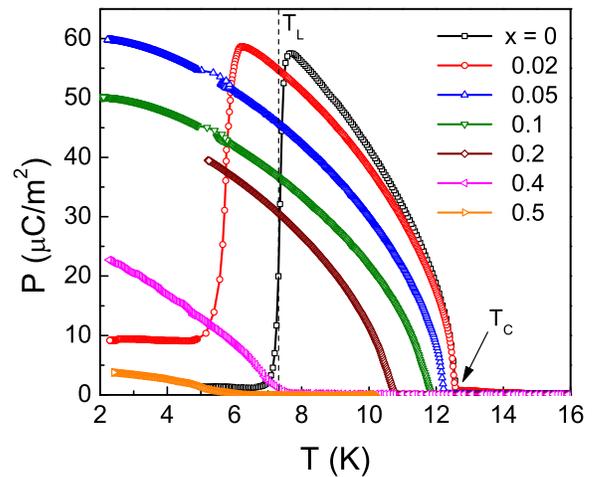}
\end{center}
\caption{(Color online) Ferroelectric polarization of Mn$_{1-x}$Zn$_x$WO$_4$ in zero magnetic field. T$_N$ and T$_L$ are labeled for the x=0 data.}
\end{figure}

\begin{figure}
\begin{center}
\includegraphics[angle=0,width=3in]{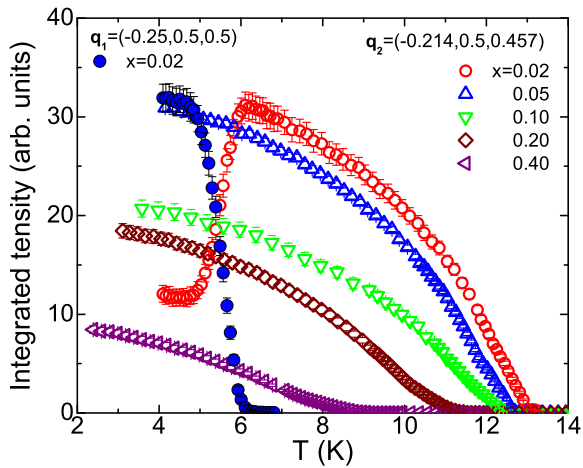}
\end{center}
\caption{(Color online) Neutron scattering intensities of the incommensurate (open symbols, $q_{2}=(-0.214,1/2,0.457)$) and commensurate (filled symbols, $q_{1}=(-0.25,0.5,0.5)$) peaks of Mn$_{1-x}$Zn$_x$WO$_4$.}
\end{figure}

At the high-temperature side the onset of magnetic order (AF3 phase) is clearly visible in the sudden increase of the neutron intensities at T$_N$. The transition from the sinusoidal AF3 phase to the helical AF2 phase at T$_C$ is more difficult to detect because both phases have the same magnetic modulation vector. A small change of slope and an anomaly in the width of the magnetic scattering peak is associated with the AF3$\rightarrow$AF2 transition. The derived critical temperatures, T$_N$ and T$_C$, are included in the phase diagram of Fig.~7 (triangles).

Similarly, the two transitions at T$_N$ and T$_C$ are also difficult to distinguish in the magnetization data of Fig.~3. The heat capacity C$_p$(T), however, shows two well-defined sharp anomalies at the two phase transitions, a sharp rise and peak at T$_N$ and a second peak at T$_C$. We have measured the heat capacity for all samples as shown in Fig. 6. Both transitions are well resolved although the width of the transitions slightly increases with x. For the undoped MnWO$_4$ (x=0) the sharp peak at 7.8 K indicates the first order phase transition into the AF1 phase. The T-x phase diagram is constructed from heat capacity, magnetization, and polarization measurements (Fig. 7). The commensurate AF1 phase is completely suppressed for x$\geq$0.05. The finite value of the low-temperature polarization for x=0.02 reveals the coexistence of AF1 and AF2 below T$_L$, as confirmed by the neutron scattering experiments (Fig. 5).

\begin{figure}
\begin{center}
\includegraphics[angle=0,width=3in]{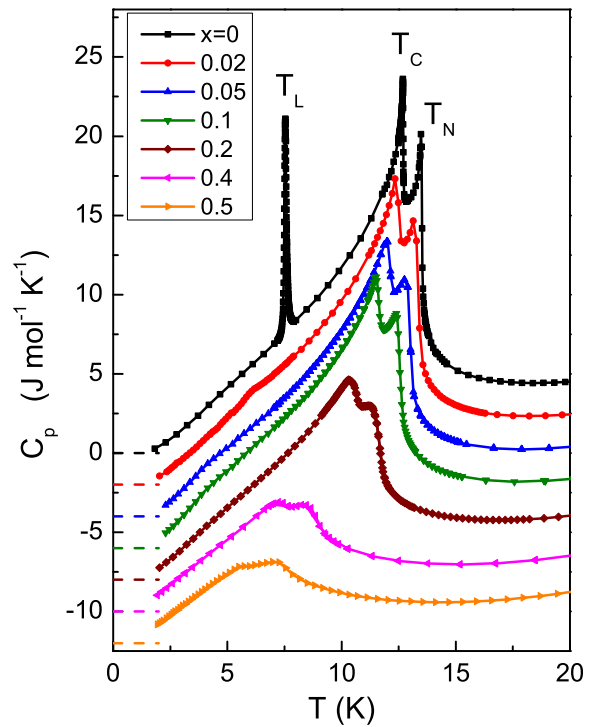}
\end{center}
\caption{(Color online) Heat capacity, C$_p$/T, of Mn$_{1-x}$Zn$_x$WO$_4$ in zero magnetic field. Different curves are vertically offset by two units (zero is defined by the dashed line for each curve).}
\end{figure}

\begin{figure}
\begin{center}
\includegraphics[angle=0,width=3in]{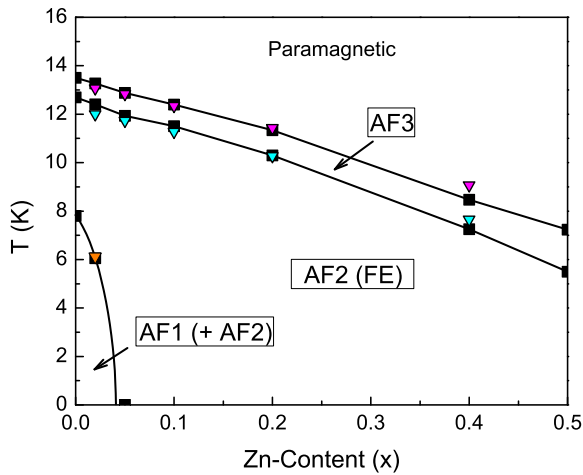}
\end{center}
\caption{(Color online) Multiferroic phase diagram of Mn$_{1-x}$Zn$_x$WO$_4$. The low-temperature phase below x=0.05 is a mixed phase, AF1 + AF2. Squares denote data from magnetic, heat capacity, and polarization measurements. Triangles are data from neutron scattering.}
\end{figure}

\begin{figure}
\begin{center}
\includegraphics[angle=0,width=3in]{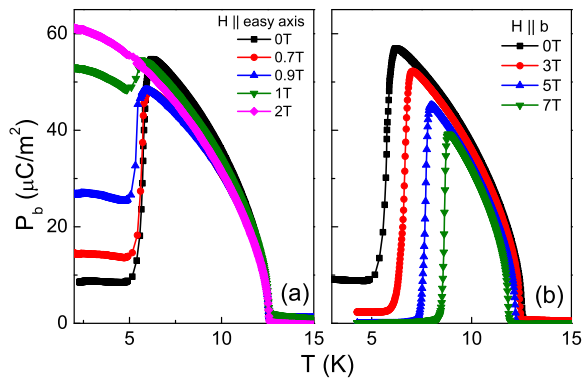}
\end{center}
\caption{(Color online) Magnetic field effect on the polarization of Mn$_{0.98}$Zn$_{0.02}$WO$_4$. The phase coexistence at low T is lifted in external fields.}
\end{figure}

The two phases (AF1 + AF2) coexisting for x$\leq$0.05 at low T can be tuned by external magnetic fields and the degeneracy is completely lifted above a critical field. This is best demonstrated in measuring the ferroelectric polarization of Mn$_{0.98}$Zn$_{0.02}$WO$_4$ in external magnetic fields with different orientations (Fig. 8). With the field directed along the spin easy axis (a-c plane, Fig. 8a) the low-T polarization (T$<$T$_L$) rapidly increases and for H$\geq$2 T the small drop of P$_b$ at T$_L$ is gone. The magnetic field has completely suppressed the AF1 phase and the multiferroic AF2 phase extends entirely to zero temperature. However, with the field applied along the b-axis (Fig. 8b) the low-T polarization decreases and reaches zero above 3 T. The low temperature phase and the ground state is now the AF1 phase and the degeneracy of both phases is removed. The lock-in transition temperature increases with H$_b$ from 5.7 K (zero field) to 8.7 K at 7 T.

\section{Summary and Conclusions}
We have shown that the multiferroic AF2 phase in Mn$_{1-x}$Zn$_x$WO$_4$ is remarkably stable with respect to the dilution of the magnetic exchange couplings induced by the substitution of nonmagnetic Zn ions. At low substitution levels and low temperatures a coexistence of two phases, the paraelectric AF1 phase and the ferroelectric AF2 phase, is shown. External magnetic fields do resolve the degeneracy of the two phases. Depending on the orientation of the magnetic field either one of the coexisting states may become the ground state.

The rapid suppression of the collinear AF1 phase by less than 5 \% Zn doping and the survival of the sinusoidal AF3 and helical AF2 phases up to 50 \% substitution is unique and distinguishes the multiferroic MnWO$_4$ from other multiferroics with a similar phase sequence, e.g. Ni$_3$V$_2$O$_8$.\cite{lawes:05} In the latter compound it was shown that the magnetic and multiferroic phases are strongly suppressed with the substitution of the magnetic Ni by nonmagnetic Zn, consistent with a two-dimensional spin system.\cite{kharel:09} In MnWO$_4$, however, the relatively small effect of the dilution of the system of Mn-spins through Zn doping on the AF3 and AF2 critical temperatures indicates a three-dimensional character of the basic magnetic exchange interactions. The microscopic exchange interactions can be revealed through inelastic neutron scattering (INS) experiments probing the magnetic excitations. By comparing the magnon spectra with standard models the most relevant exchange parameters can be extracted. According to recent INS experiments on MnWO$_4$ short range (nearest and next-nearest neighbors) exchange interactions are not sufficient to explain the magnetic excitation spectrum but up to 11 different exchange pathways have to be involved to fit the data.\cite{ehrenberg:99,ye:10} The higher order exchange coupling constants involved in the magnetic correlations prove the three-dimensional character of the magnetic fluctuations. In three dimensions the percolation threshold for site dilution is much lower than for two-dimensional systems explaining the robustness of the magnetic orders (AF3 and AF2) in Zn-doped MnWO$_4$.

The low-temperature AF1 phase, however, is in strong competition with the helical AF2 state. A simple mean-field model calculation has shown that the phase boundary between AF1 and AF2 as function of competing exchange coupling and anisotropy constants can be very steep and extremely sensitive with respect to small perturbations.\cite{kenzelmann:06,chaudhury:08} The frustration of different magnetic states results in the deviation from the mean field result for the suppression of the critical temperature as a function of doping level in a diluted magnetic system and the strong suppression of the AF1 phase as observed in Mn$_{1-x}$Zn$_x$WO$_4$.

\textit{Note added}: After submission of this manuscript the authors became aware of a related investigation of polycrystalline MnWO$_4$ doped with Mg and Zn up to 30 \%.\cite{meddar:09} The conclusions derived from magnetization and dielectric constant data are consistent with the phase diagram of Fig. 7.

\acknowledgments
This work is supported in part by the T.L.L. Temple Foundation, the J. J. and R. Moores Endowment, and the State of Texas through TCSUH and at LBNL through the US DOE, Contract No. DE-AC03-76SF00098. The research at Oak Ridge National Laboratory's High Flux Isotope Reactor was sponsored by the Scientific User Facilities, Office of Basic Energy Sciences, U. S. Department of Energy.

%\bibliographystyle{phpf}

%\bibliography{HMO}

\begin{thebibliography}{10}%
\makeatletter
\providecommand \@ifxundefined [1]{%
 \ifx #1\undefined \expandafter \@firstoftwo
 \else \expandafter \@secondoftwo
\fi
}%
\providecommand \@ifnum [1]{%
 \ifnum #1\expandafter \@firstoftwo
 \else \expandafter \@secondoftwo
\fi
}%
\providecommand \enquote [1]{``#1''}%
\providecommand \bibnamefont  [1]{#1}%
\providecommand \bibfnamefont [1]{#1}%
\providecommand \citenamefont [1]{#1}%
\providecommand\href[0]{\@sanitize\@href}%
\providecommand\@href[1]{\endgroup\@@startlink{#1}\endgroup\@@href}%
\providecommand\@@href[1]{#1\@@endlink}%
\providecommand \@sanitize [0]{\begingroup\catcode`\&12\catcode`\#12\relax}%
\@ifxundefined \pdfoutput {\@firstoftwo}{%
 \@ifnum{\z@=\pdfoutput}{\@firstoftwo}{\@secondoftwo}%
}{%
 \providecommand\@@startlink[1]{\leavevmode}%
 \providecommand\@@endlink[0]{}%
}{%
 \providecommand\@@startlink[1]{%
  \leavevmode
  \pdfstartlink
   attr{/Border[0 0 1 ]/H/I/C[0 1 1]}%
   user{/Subtype/Link/A<</Type/Action/S/URI/URI(#1)>>}%
  \relax
 }%
 \providecommand\@@endlink[0]{\pdfendlink}%
}%
\providecommand \url  [0]{\begingroup\@sanitize \@url }%
\providecommand \@url [1]{\endgroup\@href {#1}{\urlprefix}}%
\providecommand \urlprefix [0]{URL }%
\providecommand \Eprint[0]{\href }%
\@ifxundefined \urlstyle {%
  \providecommand \doi [1]{doi:\discretionary{}{}{}#1}%
}{%
  \providecommand \doi [0]{doi:\discretionary{}{}{}\begingroup
  \urlstyle{rm}\Url }%
}%
\providecommand \doibase [0]{http://dx.doi.org/}%
\providecommand \Doi[1]{\href{\doibase#1}}%
\providecommand \bibAnnote [3]{%
  \BibitemShut{#1}%
  \begin{quotation}\noindent
    \textsc{Key:}\ #2\\\textsc{Annotation:}\ #3%
  \end{quotation}%
}%
\providecommand \bibAnnoteFile [2]{%
  \IfFileExists{#2}{\bibAnnote {#1} {#2} {\input{#2}}}{}%
}%
\providecommand \typeout [0]{\immediate \write \m@ne }%
\providecommand \selectlanguage [0]{\@gobble}%
\providecommand \bibinfo [0]{\@secondoftwo}%
\providecommand \bibfield [0]{\@secondoftwo}%
\providecommand \translation [1]{[#1]}%
\providecommand \BibitemOpen[0]{}%
\providecommand \bibitemStop [0]{}%
\providecommand \bibitemNoStop [0]{.\EOS\space}%
\providecommand \EOS [0]{\spacefactor3000\relax}%
\providecommand \BibitemShut [1]{\csname bibitem#1\endcsname}%
%</preamble>
\bibitem{fiebig:05}%
  \BibitemOpen
  \bibfield{author}{%
  \bibinfo {author} {\bibfnamefont{M.}~\bibnamefont{Fiebig}},\ }%
  \bibfield{journal}{%
  \bibinfo {journal} {J. Phys. D: Appl. Phys.}\ }%
  \textbf{\bibinfo {volume} {38}},\ \bibinfo {pages} {R123} (\bibinfo {year}
  {2005})%
  \bibAnnoteFile{NoStop}{fiebig:05}%
\bibitem{spaldin:05}%
  \BibitemOpen
  \bibfield{author}{%
  \bibinfo {author} {\bibfnamefont{N.~A.}\ \bibnamefont{Spaldin}}\ and\
  \bibinfo {author} {\bibfnamefont{M.}~\bibnamefont{Fiebig}},\ }%
  \bibfield{journal}{%
  \bibinfo {journal} {Science}\ }%
  \textbf{\bibinfo {volume} {309}},\ \bibinfo {pages} {391} (\bibinfo {year}
  {2005})%
  \bibAnnoteFile{NoStop}{spaldin:05}%
\bibitem{tokura:07}%
  \BibitemOpen
  \bibfield{author}{%
  \bibinfo {author} {\bibfnamefont{Y.}~\bibnamefont{Tokura}},\ }%
  \bibfield{journal}{%
  \bibinfo {journal} {J. Mag. Mag. Mat.}\ }%
  \textbf{\bibinfo {volume} {310}},\ \bibinfo {pages} {1145} (\bibinfo {year}
  {2007})%
  \bibAnnoteFile{NoStop}{tokura:07}%
\bibitem{kenzelmann:05}%
  \BibitemOpen
  \bibfield{author}{%
  \bibinfo {author} {\bibfnamefont{M.}~\bibnamefont{Kenzelmann}}, \bibinfo
  {author} {\bibfnamefont{A.~B.}\ \bibnamefont{Harris}}, \bibinfo {author}
  {\bibfnamefont{S.}~\bibnamefont{Jonas}}, \bibinfo {author}
  {\bibfnamefont{C.}~\bibnamefont{Broholm}}, \bibinfo {author}
  {\bibfnamefont{J.}~\bibnamefont{Schefer}}, \bibinfo {author}
  {\bibfnamefont{S.~B.}\ \bibnamefont{Kim}}, \bibinfo {author}
  {\bibfnamefont{C.~L.}\ \bibnamefont{Zhang}}, \bibinfo {author}
  {\bibfnamefont{S.-W.}\ \bibnamefont{Cheong}}, \bibinfo {author}
  {\bibfnamefont{O.~P.}\ \bibnamefont{Vajk}},\ and\ \bibinfo {author}
  {\bibfnamefont{J.~W.}\ \bibnamefont{Lynn}},\ }%
  \bibfield{journal}{%
  \bibinfo {journal} {Phys. Rev. Lett.}\ }%
  \textbf{\bibinfo {volume} {95}},\ \bibinfo {pages} {087206} (\bibinfo {year}
  {2005})%
  \bibAnnoteFile{NoStop}{kenzelmann:05}%
\bibitem{mostovoy:06}%
  \BibitemOpen
  \bibfield{author}{%
  \bibinfo {author} {\bibfnamefont{M.}~\bibnamefont{Mostovoy}},\ }%
  \bibfield{journal}{%
  \bibinfo {journal} {Phys. Rev. Lett.}\ }%
  \textbf{\bibinfo {volume} {96}},\ \bibinfo {pages} {067601} (\bibinfo {year}
  {2006})%
  \bibAnnoteFile{NoStop}{mostovoy:06}%
\bibitem{lawes:05}%
  \BibitemOpen
  \bibfield{author}{%
  \bibinfo {author} {\bibfnamefont{G.}~\bibnamefont{Lawes}}, \bibinfo {author}
  {\bibfnamefont{A.~B.}\ \bibnamefont{Harris}}, \bibinfo {author}
  {\bibfnamefont{T.}~\bibnamefont{Kimura}}, \bibinfo {author}
  {\bibfnamefont{N.}~\bibnamefont{Rogado}}, \bibinfo {author}
  {\bibfnamefont{R.~J.}\ \bibnamefont{Cava}}, \bibinfo {author}
  {\bibfnamefont{A.}~\bibnamefont{Aharony}}, \bibinfo {author}
  {\bibfnamefont{O.}~\bibnamefont{Entin-Wohlman}}, \bibinfo {author}
  {\bibfnamefont{T.}~\bibnamefont{Yildirim}}, \bibinfo {author}
  {\bibfnamefont{M.}~\bibnamefont{Kenzelmann}}, \bibinfo {author}
  {\bibfnamefont{C.}~\bibnamefont{Broholm}},\ and\ \bibinfo {author}
  {\bibfnamefont{A.~P.}\ \bibnamefont{Ramirez}},\ }%
  \bibfield{journal}{%
  \bibinfo {journal} {Phys. Rev. Lett.}\ }%
  \textbf{\bibinfo {volume} {95}},\ \bibinfo {pages} {087205} (\bibinfo {year}
  {2005})%
  \bibAnnoteFile{NoStop}{lawes:05}%
\bibitem{taniguchi:06}%
  \BibitemOpen
  \bibfield{author}{%
  \bibinfo {author} {\bibfnamefont{K.}~\bibnamefont{Taniguchi}}, \bibinfo
  {author} {\bibfnamefont{N.}~\bibnamefont{Abe}}, \bibinfo {author}
  {\bibfnamefont{T.}~\bibnamefont{Takenobu}}, \bibinfo {author}
  {\bibfnamefont{Y.}~\bibnamefont{Iwasa}},\ and\ \bibinfo {author}
  {\bibfnamefont{T.}~\bibnamefont{Arima}},\ }%
  \bibfield{journal}{%
  \bibinfo {journal} {Phys. Rev. Lett.}\ }%
  \textbf{\bibinfo {volume} {97}},\ \bibinfo {pages} {097203} (\bibinfo {year}
  {2006})%
  \bibAnnoteFile{NoStop}{taniguchi:06}%
\bibitem{heyer:06}%
  \BibitemOpen
  \bibfield{author}{%
  \bibinfo {author} {\bibfnamefont{O.}~\bibnamefont{Heyer}}, \bibinfo {author}
  {\bibfnamefont{N.}~\bibnamefont{Hollmann}}, \bibinfo {author}
  {\bibfnamefont{I.}~\bibnamefont{Klassen}}, \bibinfo {author}
  {\bibfnamefont{S.}~\bibnamefont{Jodlauk}}, \bibinfo {author}
  {\bibfnamefont{L.}~\bibnamefont{Bohaty}}, \bibinfo {author}
  {\bibfnamefont{P.}~\bibnamefont{Becker}}, \bibinfo {author}
  {\bibfnamefont{J.~A.}\ \bibnamefont{Mydosh}}, \bibinfo {author}
  {\bibfnamefont{T.}~\bibnamefont{Lorenz}},\ and\ \bibinfo {author}
  {\bibfnamefont{D.}~\bibnamefont{Khomskii}},\ }%
  \bibfield{journal}{%
  \bibinfo {journal} {J. Phys.: Condens. Matter}\ }%
  \textbf{\bibinfo {volume} {18}},\ \bibinfo {pages} {L471} (\bibinfo {year}
  {2006})%
  \bibAnnoteFile{NoStop}{heyer:06}%
\bibitem{katsura:05}%
  \BibitemOpen
  \bibfield{author}{%
  \bibinfo {author} {\bibfnamefont{H.}~\bibnamefont{Katsura}}, \bibinfo
  {author} {\bibfnamefont{N.}~\bibnamefont{Nagaosa}},\ and\ \bibinfo {author}
  {\bibfnamefont{A.~V.}\ \bibnamefont{Balatsky}},\ }%
  \bibfield{journal}{%
  \bibinfo {journal} {Phys. Rev. Lett.}\ }%
  \textbf{\bibinfo {volume} {95}},\ \bibinfo {pages} {057205} (\bibinfo {year}
  {2005})%
  \bibAnnoteFile{NoStop}{katsura:05}%
\bibitem{sergienko:06}%
  \BibitemOpen
  \bibfield{author}{%
  \bibinfo {author} {\bibfnamefont{I.~A.}\ \bibnamefont{Sergienko}}\ and\
  \bibinfo {author} {\bibfnamefont{E.}~\bibnamefont{Dagotto}},\ }%
  \bibfield{journal}{%
  \bibinfo {journal} {Phys. Rev. B}\ }%
  \textbf{\bibinfo {volume} {73}},\ \bibinfo {pages} {094434} (\bibinfo {year}
  {2006})%
  \bibAnnoteFile{NoStop}{sergienko:06}%
\bibitem{sergienko:06b}%
  \BibitemOpen
  \bibfield{author}{%
  \bibinfo {author} {\bibfnamefont{I.~A.}\ \bibnamefont{Sergienko}}, \bibinfo
  {author} {\bibfnamefont{C.}~\bibnamefont{Sen}},\ and\ \bibinfo {author}
  {\bibfnamefont{E.}~\bibnamefont{Dagotto}},\ }%
  \bibfield{journal}{%
  \bibinfo {journal} {Phys. Rev. Lett.}\ }%
  \textbf{\bibinfo {volume} {97}},\ \bibinfo {pages} {227204} (\bibinfo {year}
  {2006})%
  \bibAnnoteFile{NoStop}{sergienko:06b}%
\bibitem{mochiguzi:10}%
  \BibitemOpen
  \bibfield{author}{%
  \bibinfo {author} {\bibfnamefont{M.}~\bibnamefont{Mochizuki}}, \bibinfo
  {author} {\bibfnamefont{N.}~\bibnamefont{Furukawa}},\ and\ \bibinfo {author}
  {\bibfnamefont{N.}~\bibnamefont{Nagaosa}},\ }%
  \bibfield{journal}{%
  \bibinfo {journal} {Phys. Rev. Lett.}\ }%
  \textbf{\bibinfo {volume} {105}},\ \bibinfo {pages} {037205} (\bibinfo {year}
  {2010})%
  \bibAnnoteFile{NoStop}{mochiguzi:10}%
\bibitem{higashiyama:04}%
  \BibitemOpen
  \bibfield{author}{%
  \bibinfo {author} {\bibfnamefont{D.}~\bibnamefont{Higashiyama}}, \bibinfo
  {author} {\bibfnamefont{S.}~\bibnamefont{Miyasaka}}, \bibinfo {author}
  {\bibfnamefont{N.}~\bibnamefont{Kida}}, \bibinfo {author}
  {\bibfnamefont{T.}~\bibnamefont{Arima}},\ and\ \bibinfo {author}
  {\bibfnamefont{Y.}~\bibnamefont{Tokura}},\ }%
  \bibfield{journal}{%
  \bibinfo {journal} {Phys. Rev. B}\ }%
  \textbf{\bibinfo {volume} {70}},\ \bibinfo {pages} {174405} (\bibinfo {year}
  {2004})%
  \bibAnnoteFile{NoStop}{higashiyama:04}%
\bibitem{hur:04}%
  \BibitemOpen
  \bibfield{author}{%
  \bibinfo {author} {\bibfnamefont{N.}~\bibnamefont{Hur}}, \bibinfo {author}
  {\bibfnamefont{S.}~\bibnamefont{Park}}, \bibinfo {author}
  {\bibfnamefont{P.~A.}\ \bibnamefont{Sharma}}, \bibinfo {author}
  {\bibfnamefont{S.}~\bibnamefont{Guha}},\ and\ \bibinfo {author}
  {\bibfnamefont{S.-W.}\ \bibnamefont{Cheong}},\ }%
  \bibfield{journal}{%
  \bibinfo {journal} {Phys. Rev. Lett.}\ }%
  \textbf{\bibinfo {volume} {93}},\ \bibinfo {pages} {107207} (\bibinfo {year}
  {2004})%
  \bibAnnoteFile{NoStop}{hur:04}%
\bibitem{seki:08}%
  \BibitemOpen
  \bibfield{author}{%
  \bibinfo {author} {\bibfnamefont{S.}~\bibnamefont{Seki}}, \bibinfo {author}
  {\bibfnamefont{Y.}~\bibnamefont{Yamasaki}}, \bibinfo {author}
  {\bibfnamefont{M.}~\bibnamefont{Soda}}, \bibinfo {author}
  {\bibfnamefont{M.}~\bibnamefont{Matsuura}}, \bibinfo {author}
  {\bibfnamefont{K.}~\bibnamefont{Hirota}},\ and\ \bibinfo {author}
  {\bibfnamefont{Y.}~\bibnamefont{Tokura}},\ }%
  \bibfield{journal}{%
  \bibinfo {journal} {Phys. Rev. Lett.}\ }%
  \textbf{\bibinfo {volume} {100}},\ \bibinfo {pages} {127201} (\bibinfo {year}
  {2008})%
  \bibAnnoteFile{NoStop}{seki:08}%
\bibitem{delacruz:07}%
  \BibitemOpen
  \bibfield{author}{%
  \bibinfo {author} {\bibfnamefont{C.~R.}\ \bibnamefont{dela Cruz}}, \bibinfo
  {author} {\bibfnamefont{B.}~\bibnamefont{Lorenz}}, \bibinfo {author}
  {\bibfnamefont{Y.~Y.}\ \bibnamefont{Sun}}, \bibinfo {author}
  {\bibfnamefont{Y.}~\bibnamefont{Wang}}, \bibinfo {author}
  {\bibfnamefont{S.}~\bibnamefont{Park}}, \bibinfo {author}
  {\bibfnamefont{S.-W.}\ \bibnamefont{Cheong}}, \bibinfo {author}
  {\bibfnamefont{M.~M.}\ \bibnamefont{Gospodinov}},\ and\ \bibinfo {author}
  {\bibfnamefont{C.~W.}\ \bibnamefont{Chu}},\ }%
  \bibfield{journal}{%
  \bibinfo {journal} {Phys. Rev. B}\ }%
  \textbf{\bibinfo {volume} {76}},\ \bibinfo {pages} {174106} (\bibinfo {year}
  {2007})%
  \bibAnnoteFile{NoStop}{delacruz:07}%
\bibitem{delacruz:08}%
  \BibitemOpen
  \bibfield{author}{%
  \bibinfo {author} {\bibfnamefont{C.~R.}\ \bibnamefont{dela Cruz}}, \bibinfo
  {author} {\bibfnamefont{B.}~\bibnamefont{Lorenz}},\ and\ \bibinfo {author}
  {\bibfnamefont{C.~W.}\ \bibnamefont{Chu}},\ }%
  \bibfield{journal}{%
  \bibinfo {journal} {Physica B}\ }%
  \textbf{\bibinfo {volume} {403}},\ \bibinfo {pages} {1331} (\bibinfo {year}
  {2008})%
  \bibAnnoteFile{NoStop}{delacruz:08}%
\bibitem{chaudhury:07}%
  \BibitemOpen
  \bibfield{author}{%
  \bibinfo {author} {\bibfnamefont{R.~P.}\ \bibnamefont{Chaudhury}}, \bibinfo
  {author} {\bibfnamefont{F.}~\bibnamefont{Yen}}, \bibinfo {author}
  {\bibfnamefont{C.~R.}\ \bibnamefont{dela Cruz}}, \bibinfo {author}
  {\bibfnamefont{B.}~\bibnamefont{Lorenz}}, \bibinfo {author}
  {\bibfnamefont{Y.~Q.}\ \bibnamefont{Wang}}, \bibinfo {author}
  {\bibfnamefont{Y.~Y.}\ \bibnamefont{Sun}},\ and\ \bibinfo {author}
  {\bibfnamefont{C.~W.}\ \bibnamefont{Chu}},\ }%
  \bibfield{journal}{%
  \bibinfo {journal} {Phys. Rev. B}\ }%
  \textbf{\bibinfo {volume} {75}},\ \bibinfo {pages} {012407} (\bibinfo {year}
  {2007})%
  \bibAnnoteFile{NoStop}{chaudhury:07}%
\bibitem{chaudhury:08b}%
  \BibitemOpen
  \bibfield{author}{%
  \bibinfo {author} {\bibfnamefont{R.~P.}\ \bibnamefont{Chaudhury}}, \bibinfo
  {author} {\bibfnamefont{C.~R.}\ \bibnamefont{dela Cruz}}, \bibinfo {author}
  {\bibfnamefont{B.}~\bibnamefont{Lorenz}}, \bibinfo {author}
  {\bibfnamefont{Y.}\ \bibnamefont{Sun}}, \bibinfo {author}
  {\bibfnamefont{C.~W.}\ \bibnamefont{Chu}}, \bibinfo {author}
  {\bibfnamefont{S.}~\bibnamefont{Park}},\ and\ \bibinfo {author}
  {\bibfnamefont{Sang-W.}\ \bibnamefont{Cheong}},\ }%
  \bibfield{journal}{%
  \bibinfo {journal} {Phys. Rev. B}\ }%
  \textbf{\bibinfo {volume} {77}},\ \bibinfo {pages} {220104(R)} (\bibinfo
  {year} {2008})%
  \bibAnnoteFile{NoStop}{chaudhury:08b}%
\bibitem{chaudhury:08}%
  \BibitemOpen
  \bibfield{author}{%
  \bibinfo {author} {\bibfnamefont{R.~P.}\ \bibnamefont{Chaudhury}}, \bibinfo
  {author} {\bibfnamefont{B.}~\bibnamefont{Lorenz}}, \bibinfo {author}
  {\bibfnamefont{Y.~Q.}\ \bibnamefont{Wang}}, \bibinfo {author}
  {\bibfnamefont{Y.~Y.}\ \bibnamefont{Sun}},\ and\ \bibinfo {author}
  {\bibfnamefont{C.~W.}\ \bibnamefont{Chu}},\ }%
  \bibfield{journal}{%
  \bibinfo {journal} {Phys. Rev. B}\ }%
  \textbf{\bibinfo {volume} {77}},\ \bibinfo {pages} {104406} (\bibinfo {year}
  {2008})%
  \bibAnnoteFile{NoStop}{chaudhury:08}%
\bibitem{chaudhury:09b}%
  \BibitemOpen
  \bibfield{author}{%
  \bibinfo {author} {\bibfnamefont{R.~P.}\ \bibnamefont{Chaudhury}}, \bibinfo
  {author} {\bibfnamefont{B.}~\bibnamefont{Lorenz}}, \bibinfo {author}
  {\bibfnamefont{Y.-Q.}\ \bibnamefont{Wang}}, \bibinfo {author}
  {\bibfnamefont{Y.~Y.}\ \bibnamefont{Sun}},\ and\ \bibinfo {author}
  {\bibfnamefont{C.~W.}\ \bibnamefont{Chu}},\ }%
  \bibfield{journal}{%
  \bibinfo {journal} {New J. Phys.}\ }%
  \textbf{\bibinfo {volume} {11}},\ \bibinfo {pages} {033036} (\bibinfo {year}
  {2009})%
  \bibAnnoteFile{NoStop}{chaudhury:09b}%
\bibitem{seki:07}%
  \BibitemOpen
  \bibfield{author}{%
  \bibinfo {author} {\bibfnamefont{S.}~\bibnamefont{Seki}}, \bibinfo {author}
  {\bibfnamefont{Y.}~\bibnamefont{Yamasaki}}, \bibinfo {author}
  {\bibfnamefont{Y.}~\bibnamefont{Shiomi}}, \bibinfo {author}
  {\bibfnamefont{S.}~\bibnamefont{Iguchi}}, \bibinfo {author}
  {\bibfnamefont{Y.}~\bibnamefont{Onose}},\ and\ \bibinfo {author}
  {\bibfnamefont{Y.}~\bibnamefont{Tokura}},\ }%
  \bibfield{journal}{%
  \bibinfo {journal} {Phys. Rev. B}\ }%
  \textbf{\bibinfo {volume} {75}},\ \bibinfo {pages} {100403(R)} (\bibinfo
  {year} {2007})%
  \bibAnnoteFile{NoStop}{seki:07}%
\bibitem{kanetsuki:07}%
  \BibitemOpen
  \bibfield{author}{%
  \bibinfo {author} {\bibfnamefont{S.}~\bibnamefont{Kanetsuki}}, \bibinfo
  {author} {\bibfnamefont{S.}~\bibnamefont{Mitsuda}}, \bibinfo {author}
  {\bibfnamefont{T.}~\bibnamefont{Nakajima}}, \bibinfo {author}
  {\bibfnamefont{D.}~\bibnamefont{Anazawa}}, \bibinfo {author}
  {\bibfnamefont{H.~A.}\ \bibnamefont{Katori}},\ and\ \bibinfo {author}
  {\bibfnamefont{K.}~\bibnamefont{Prokes}},\ }%
  \bibfield{journal}{%
  \bibinfo {journal} {J. Phys.: Condens. Matter}\ }%
  \textbf{\bibinfo {volume} {19}},\ \bibinfo {pages} {145244} (\bibinfo {year}
  {2007})%
  \bibAnnoteFile{NoStop}{kanetsuki:07}%
\bibitem{lautenschlager:93}%
  \BibitemOpen
  \bibfield{author}{%
  \bibinfo {author} {\bibfnamefont{G.}~\bibnamefont{Lautenschl{\"a}ger}},
  \bibinfo {author} {\bibfnamefont{H.}~\bibnamefont{Weitzel}}, \bibinfo
  {author} {\bibfnamefont{T.}~\bibnamefont{Vogt}}, \bibinfo {author}
  {\bibfnamefont{R.}~\bibnamefont{Hock}}, \bibinfo {author}
  {\bibfnamefont{A.}~\bibnamefont{Bohm}}, \bibinfo {author}
  {\bibfnamefont{M.}~\bibnamefont{Bonnet}},\ and\ \bibinfo {author}
  {\bibfnamefont{H.}~\bibnamefont{Fuess}},\ }%
  \bibfield{journal}{%
  \bibinfo {journal} {Phys. Rev. B}\ }%
  \textbf{\bibinfo {volume} {48}},\ \bibinfo {pages} {6087} (\bibinfo {year}
  {1993})%
  \bibAnnoteFile{NoStop}{lautenschlager:93}%
\bibitem{sagayama:08}%
  \BibitemOpen
  \bibfield{author}{%
  \bibinfo {author} {\bibfnamefont{H.}~\bibnamefont{Sagayama}}, \bibinfo
  {author} {\bibfnamefont{K.}~\bibnamefont{Taniguchi}}, \bibinfo {author}
  {\bibfnamefont{N.}~\bibnamefont{Abe}}, \bibinfo {author}
  {\bibfnamefont{T.~H.}~\bibnamefont{Arima}}, \bibinfo {author}
  {\bibfnamefont{M.}~\bibnamefont{Soda}}, \bibinfo {author}
  {\bibfnamefont{M.}~\bibnamefont{Matsuura}},\ and\ \bibinfo {author}
  {\bibfnamefont{K.}~\bibnamefont{Hirota}},\ }%
  \bibfield{journal}{%
  \bibinfo {journal} {Phys. Rev. B}\ }%
  \textbf{\bibinfo {volume} {77}},\ \bibinfo {pages} {220407(R)} (\bibinfo
  {year} {2008})%
  \bibAnnoteFile{NoStop}{sagayama:08}%
\bibitem{ehrenberg:97}%
  \BibitemOpen
  \bibfield{author}{%
  \bibinfo {author} {\bibfnamefont{H.}~\bibnamefont{Ehrenberg}}, \bibinfo
  {author} {\bibfnamefont{H.}~\bibnamefont{Weitzel}}, \bibinfo {author}
  {\bibfnamefont{C.}~\bibnamefont{Heidy}}, \bibinfo {author}
  {\bibfnamefont{H.}~\bibnamefont{Fuess}}, \bibinfo {author}
  {\bibfnamefont{G.}~\bibnamefont{Wltschek}}, \bibinfo {author}
  {\bibfnamefont{T.}~\bibnamefont{Kroener}}, \bibinfo {author}
  {\bibfnamefont{J.}~\bibnamefont{van Tol}},\ and\ \bibinfo {author}
  {\bibfnamefont{M.}~\bibnamefont{Bonner}},\ }%
  \bibfield{journal}{%
  \bibinfo {journal} {J. Phys. Condens. Matter}\ }%
  \textbf{\bibinfo {volume} {9}},\ \bibinfo {pages} {3189} (\bibinfo {year}
  {1997})%
  \bibAnnoteFile{NoStop}{ehrenberg:97}%
\bibitem{arkenbout:06}%
  \BibitemOpen
  \bibfield{author}{%
  \bibinfo {author} {\bibfnamefont{A.~H.}\ \bibnamefont{Arkenbout}}, \bibinfo
  {author} {\bibfnamefont{T.~T.~M.}\ \bibnamefont{Palstra}}, \bibinfo {author}
  {\bibfnamefont{T.}~\bibnamefont{Siegrist}},\ and\ \bibinfo {author}
  {\bibfnamefont{T.}~\bibnamefont{Kimura}},\ }%
  \bibfield{journal}{%
  \bibinfo {journal} {Phys. Rev. B}\ }%
  \textbf{\bibinfo {volume} {74}},\ \bibinfo {pages} {184431} (\bibinfo {year}
  {2006})%
  \bibAnnoteFile{NoStop}{arkenbout:06}%
\bibitem{taniguchi:08}%
  \BibitemOpen
  \bibfield{author}{%
  \bibinfo {author} {\bibfnamefont{K.}~\bibnamefont{Taniguchi}}, \bibinfo
  {author} {\bibfnamefont{N.}~\bibnamefont{Abe}}, \bibinfo {author}
  {\bibfnamefont{H.}~\bibnamefont{Sagayama}}, \bibinfo {author}
  {\bibfnamefont{S.}~\bibnamefont{Ohtani}}, \bibinfo {author}
  {\bibfnamefont{T.}~\bibnamefont{Takenobu}}, \bibinfo {author}
  {\bibfnamefont{Y.}~\bibnamefont{Iwasa}},\ and\ \bibinfo {author}
  {\bibfnamefont{T.}~\bibnamefont{Arima}},\ }%
  \bibfield{journal}{%
  \bibinfo {journal} {Phys. Rev. B}\ }%
  \textbf{\bibinfo {volume} {77}},\ \bibinfo {pages} {064408} (\bibinfo {year}
  {2008})%
  \bibAnnoteFile{NoStop}{taniguchi:08}%
\bibitem{chaudhury:08c}%
  \BibitemOpen
  \bibfield{author}{%
  \bibinfo {author} {\bibfnamefont{R.~P.}\ \bibnamefont{Chaudhury}}, \bibinfo
  {author} {\bibfnamefont{F.}~\bibnamefont{Yen}}, \bibinfo {author}
  {\bibfnamefont{C.~R.}\ \bibnamefont{dela Cruz}}, \bibinfo {author}
  {\bibfnamefont{B.}~\bibnamefont{Lorenz}}, \bibinfo {author}
  {\bibfnamefont{Y.~Q.}\ \bibnamefont{Wang}}, \bibinfo {author}
  {\bibfnamefont{Y.~Y.}\ \bibnamefont{Sun}},\ and\ \bibinfo {author}
  {\bibfnamefont{C.~W.}\ \bibnamefont{Chu}},\ }%
  \bibfield{journal}{%
  \bibinfo {journal} {Physica B}\ }%
  \textbf{\bibinfo {volume} {403}},\ \bibinfo {pages} {1428} (\bibinfo {year}
  {2008})%
  \bibAnnoteFile{NoStop}{chaudhury:08c}%
\bibitem{meier:09}%
  \BibitemOpen
  \bibfield{author}{%
  \bibinfo {author} {\bibfnamefont{D.}~\bibnamefont{Meier}}, \bibinfo {author}
  {\bibfnamefont{M.}~\bibnamefont{Maringer}}, \bibinfo {author}
  {\bibfnamefont{T.}~\bibnamefont{Lottermoser}}, \bibinfo {author}
  {\bibfnamefont{P.}~\bibnamefont{Becker}}, \bibinfo {author}
  {\bibfnamefont{L.}~\bibnamefont{Bohaty}},\ and\ \bibinfo {author}
  {\bibfnamefont{M.}~\bibnamefont{Fiebig}},\ }%
  \bibfield{journal}{%
  \bibinfo {journal} {Phys. Rev. Lett.}\ }%
  \textbf{\bibinfo {volume} {102}},\ \bibinfo {pages} {107202} (\bibinfo {year}
  {2009})%
  \bibAnnoteFile{NoStop}{meier:09}%
\bibitem{finger:10}%
  \BibitemOpen
  \bibfield{author}{%
  \bibinfo {author} {\bibfnamefont{T.}~\bibnamefont{Finger}}, \bibinfo {author}
  {\bibfnamefont{D.}~\bibnamefont{Senff}}, \bibinfo {author}
  {\bibfnamefont{K.}~\bibnamefont{Schmalzl}}, \bibinfo {author}
  {\bibfnamefont{W.}~\bibnamefont{Schmidt}}, \bibinfo {author}
  {\bibfnamefont{L.~P.}\ \bibnamefont{Regnault}}, \bibinfo {author}
  {\bibfnamefont{P.}~\bibnamefont{Becker}}, \bibinfo {author}
  {\bibfnamefont{L.}~\bibnamefont{Bohaty}},\ and\ \bibinfo {author}
  {\bibfnamefont{M.}~\bibnamefont{Braden}},\ }%
  \bibfield{journal}{%
  \bibinfo {journal} {Phys. Rev. B}\ }%
  \textbf{\bibinfo {volume} {81}},\ \bibinfo {pages} {054430} (\bibinfo {year}
  {2010})%
  \bibAnnoteFile{NoStop}{finger:10}%
\bibitem{obermayer:73}%
  \BibitemOpen
  \bibfield{author}{%
  \bibinfo {author} {\bibfnamefont{H.~A.}\ \bibnamefont{Obermayer}}, \bibinfo
  {author} {\bibfnamefont{H.}~\bibnamefont{Dachs}},\ and\ \bibinfo {author}
  {\bibfnamefont{H.}~\bibnamefont{Schr{\"o}cke}},\ }%
  \bibfield{journal}{%
  \bibinfo {journal} {Solid State Commun.}\ }%
  \textbf{\bibinfo {volume} {12}},\ \bibinfo {pages} {779} (\bibinfo {year}
  {1973})%
  \bibAnnoteFile{NoStop}{obermayer:73}%
\bibitem{klein:74}%
  \BibitemOpen
  \bibfield{author}{%
  \bibinfo {author} {\bibfnamefont{C.}~\bibnamefont{Klein}}\ and\ \bibinfo
  {author} {\bibfnamefont{R.}~\bibnamefont{Geller}},\ }%
  \bibfield{journal}{%
  \bibinfo {journal} {J. de Physique}\ }%
  \textbf{\bibinfo {volume} {35}},\ \bibinfo {pages} {C6:589} (\bibinfo {year}
  {1974})%
  \bibAnnoteFile{NoStop}{klein:74}%
\bibitem{garciamatres:03}%
  \BibitemOpen
  \bibfield{author}{%
  \bibinfo {author} {\bibfnamefont{E.}~\bibnamefont{Garcia-Matres}}, \bibinfo
  {author} {\bibfnamefont{N.}~\bibnamefont{St{\"u}${\ss}$er}}, \bibinfo
  {author} {\bibfnamefont{M.}~\bibnamefont{Hofmann}},\ and\ \bibinfo {author}
  {\bibfnamefont{M.}~\bibnamefont{Reehuis}},\ }%
  \bibfield{journal}{%
  \bibinfo {journal} {Eur. Phys. J. B}\ }%
  \textbf{\bibinfo {volume} {32}},\ \bibinfo {pages} {35} (\bibinfo {year}
  {2003})%
  \bibAnnoteFile{NoStop}{garciamatres:03}%
\bibitem{ding:00}%
  \BibitemOpen
  \bibfield{author}{%
  \bibinfo {author} {\bibfnamefont{Y.}~\bibnamefont{Ding}}, \bibinfo {author}
  {\bibfnamefont{N.}~\bibnamefont{St{\"u}${\ss}$er}}, \bibinfo {author}
  {\bibfnamefont{M.}~\bibnamefont{Reehuis}}, \bibinfo {author}
  {\bibfnamefont{M.}~\bibnamefont{Hofmann}}, \bibinfo {author}
  {\bibfnamefont{G.}~\bibnamefont{Ehlers}}, \bibinfo {author}
  {\bibfnamefont{D.}~\bibnamefont{G{\"u}nther}}, \bibinfo {author}
  {\bibfnamefont{M.}~\bibnamefont{Mei${\ss}$ner}}, \bibinfo {author}
  {\bibfnamefont{S.}~\bibnamefont{Welzel}}, \bibinfo {author}
  {\bibfnamefont{M.}~\bibnamefont{Wilhelm}}, \bibinfo {author}
  {\bibfnamefont{M.}~\bibnamefont{Steiner}},\ and\ \bibinfo {author}
  {\bibfnamefont{F.}~\bibnamefont{Kubanek}},\ }%
  \bibfield{journal}{%
  \bibinfo {journal} {Physica B}\ }%
  \textbf{\bibinfo {volume} {276-278}},\ \bibinfo {pages} {596} (\bibinfo
  {year} {2000})%
  \bibAnnoteFile{NoStop}{ding:00}%
\bibitem{song:09}%
  \BibitemOpen
  \bibfield{author}{%
  \bibinfo {author} {\bibfnamefont{Y.-S.}\ \bibnamefont{Song}}, \bibinfo
  {author} {\bibfnamefont{J.-H.}\ \bibnamefont{Chung}}, \bibinfo {author}
  {\bibfnamefont{J.~M.~S.}\ \bibnamefont{Park}},\ and\ \bibinfo {author}
  {\bibfnamefont{Y.-N.}\ \bibnamefont{Choi}},\ }%
  \bibfield{journal}{%
  \bibinfo {journal} {Phys. Rev. B}\ }%
  \textbf{\bibinfo {volume} {79}},\ \bibinfo {pages} {224415} (\bibinfo {year}
  {2009})%
  \bibAnnoteFile{NoStop}{song:09}%
\bibitem{ye:08}%
  \BibitemOpen
  \bibfield{author}{%
  \bibinfo {author} {\bibfnamefont{F.}~\bibnamefont{Ye}}, \bibinfo {author}
  {\bibfnamefont{Y.}~\bibnamefont{Ren}}, \bibinfo {author}
  {\bibfnamefont{J.~A.}\ \bibnamefont{Fernandez-Baca}}, \bibinfo {author}
  {\bibfnamefont{H.~A.}\ \bibnamefont{Mook}}, \bibinfo {author}
  {\bibfnamefont{J.~W.}\ \bibnamefont{Lynn}}, \bibinfo {author}
  {\bibfnamefont{R.~P.}\ \bibnamefont{Chaudhury}}, \bibinfo {author}
  {\bibfnamefont{Y.-Q.}\ \bibnamefont{Wang}}, \bibinfo {author}
  {\bibfnamefont{B.}~\bibnamefont{Lorenz}},\ and\ \bibinfo {author}
  {\bibfnamefont{C.~W.}\ \bibnamefont{Chu}},\ }%
  \bibfield{journal}{%
  \bibinfo {journal} {Phys. Rev. B}\ }%
  \textbf{\bibinfo {volume} {78}},\ \bibinfo {pages} {193101} (\bibinfo {year}
  {2008})%
  \bibAnnoteFile{NoStop}{ye:08}%
\bibitem{lawes:04}%
  \BibitemOpen
  \bibfield{author}{%
  \bibinfo {author} {\bibfnamefont{G.}~\bibnamefont{Lawes}}, \bibinfo {author}
  {\bibfnamefont{M.}~\bibnamefont{Kenzelmann}}, \bibinfo {author}
  {\bibfnamefont{N.}~\bibnamefont{Rogado}}, \bibinfo {author}
  {\bibfnamefont{K.~H.}\ \bibnamefont{Kim}}, \bibinfo {author}
  {\bibfnamefont{G.~A.}\ \bibnamefont{Jorge}}, \bibinfo {author}
  {\bibfnamefont{R.~J.}\ \bibnamefont{Cava}}, \bibinfo {author}
  {\bibfnamefont{A.}~\bibnamefont{Aharony}}, \bibinfo {author}
  {\bibfnamefont{O.}~\bibnamefont{Entin-Wohlman}}, \bibinfo {author}
  {\bibfnamefont{A.~B.}\ \bibnamefont{Harris}}, \bibinfo {author}
  {\bibfnamefont{T.}~\bibnamefont{Yildirim}}, \bibinfo {author}
  {\bibfnamefont{Q.~Z.}\ \bibnamefont{Huang}}, \bibinfo {author}
  {\bibfnamefont{S.}~\bibnamefont{Park}}, \bibinfo {author}
  {\bibfnamefont{C.}~\bibnamefont{Broholm}},\ and\ \bibinfo {author}
  {\bibfnamefont{A.~P.}\ \bibnamefont{Ramirez}},\ }%
  \bibfield{journal}{%
  \bibinfo {journal} {Phys. Rev. Lett.}\ }%
  \textbf{\bibinfo {volume} {93}},\ \bibinfo {pages} {247201} (\bibinfo {year}
  {2004})%
  \bibAnnoteFile{NoStop}{lawes:04}%
\bibitem{chaudhury:10}%
  \BibitemOpen
  \bibfield{author}{%
  \bibinfo {author} {\bibfnamefont{R.~P.}\ \bibnamefont{Chaudhury}}, \bibinfo
  {author} {\bibfnamefont{F.}~\bibnamefont{Ye}}, \bibinfo {author}
  {\bibfnamefont{J.~A.}\ \bibnamefont{Fernandez-Baca}}, \bibinfo {author}
  {\bibfnamefont{Y.~Q.}\ \bibnamefont{Wang}}, \bibinfo {author}
  {\bibfnamefont{Y.~Y.}\ \bibnamefont{Sun}}, \bibinfo {author}
  {\bibfnamefont{B.}~\bibnamefont{Lorenz}}, \bibinfo {author}
  {\bibfnamefont{H.~A.}~\bibnamefont{Mook}}, \ and\ \bibinfo {author}
  {\bibfnamefont{C.~W.}\ \bibnamefont{Chu}},\ }%
  \bibfield{journal}{%
  \bibinfo {journal} {Phys. Rev. B}\ }%
  \textbf{\bibinfo {volume} {82}},\ \bibinfo {pages} {184422} (\bibinfo {year}
  {2010})%
  \bibAnnoteFile{NoStop}{chaudhury:10}%
\bibitem{chaudhury:09c}%
  \BibitemOpen
  \bibfield{author}{%
  \bibinfo {author} {\bibfnamefont{R.~P.}\ \bibnamefont{Chaudhury}}, \bibinfo
  {author} {\bibfnamefont{B.}~\bibnamefont{Lorenz}}, \bibinfo {author}
  {\bibfnamefont{Y.~Q.}\ \bibnamefont{Wang}}, \bibinfo {author}
  {\bibfnamefont{Y.~Y.}\ \bibnamefont{Sun}}, \bibinfo {author}
  {\bibfnamefont{C.~W.}\ \bibnamefont{Chu}}, \bibinfo {author}
  {\bibfnamefont{F.}~\bibnamefont{Ye}}, \bibinfo {author}
  {\bibfnamefont{J.}~\bibnamefont{Fernandez-Baca}}, \bibinfo {author}
  {\bibfnamefont{H.}~\bibnamefont{Mook}}, \ and\ \bibinfo {author}
  {\bibfnamefont{J.}\ \bibnamefont{Lynn}},\ }%
  \bibfield{journal}{%
  \bibinfo {journal} {J. Appl. Phys.}\ }%
  \textbf{\bibinfo {volume} {105}},\ \bibinfo {pages} {07D913} (\bibinfo {year}
  {2009})%
  \bibAnnoteFile{NoStop}{chaudhury:09c}%
\bibitem{kharel:09}%
  \BibitemOpen
  \bibfield{author}{%
  \bibinfo {author} {\bibfnamefont{P.}\ \bibnamefont{Kharel}}, \bibinfo
  {author} {\bibfnamefont{A.}~\bibnamefont{Kumarasiri}}, \bibinfo {author}
  {\bibfnamefont{A.}\ \bibnamefont{Dixit}}, \bibinfo {author}
  {\bibfnamefont{N.}\ \bibnamefont{Rogado}}, \bibinfo {author}
  {\bibfnamefont{R.~J.}\ \bibnamefont{Cava}}, \ and\ \bibinfo {author}
  {\bibfnamefont{G.}~\bibnamefont{Lawes}},\ }%
  \bibfield{journal}{%
  \bibinfo {journal} {Phil. Mag.}\ }%
  \textbf{\bibinfo {volume} {89}},\ \bibinfo {pages} {1923} (\bibinfo {year}
  {2009})%
  \bibAnnoteFile{NoStop}{kharel:09}%
\bibitem{ehrenberg:99}%
  \BibitemOpen
  \bibfield{author}{%
  \bibinfo {author} {\bibfnamefont{H.}\ \bibnamefont{Ehrenberg}}, \bibinfo
  {author} {\bibfnamefont{H.}~\bibnamefont{Weitzel}}, \bibinfo {author}
  {\bibfnamefont{H.}\ \bibnamefont{Fuess}},\ and\ \bibinfo {author}
  {\bibfnamefont{B.}~\bibnamefont{Hennion}},\ }%
  \bibfield{journal}{%
  \bibinfo {journal} {J. Phys.: Condens. Matter}\ }%
  \textbf{\bibinfo {volume} {11}},\ \bibinfo {pages} {2649} (\bibinfo {year}
  {1999})%
  \bibAnnoteFile{NoStop}{ehrenberg:99}%
\bibitem{ye:10}%
  \BibitemOpen
  \bibfield{author}{%
  \bibinfo {author} {\bibfnamefont{F.}\ \bibnamefont{Ye}}, \bibinfo {author} {\bibfnamefont{R. S.}\ \bibnamefont{Fishman}}, \bibinfo {author} {\bibfnamefont{J. A.}\ \bibnamefont{Fernandez-Baca}}, \bibinfo {author} {\bibfnamefont{A. A.}\ \bibnamefont{Podlesnyak}}, \bibinfo {author} {\bibfnamefont{G.}\ \bibnamefont{Ehlers}}, \bibinfo {author} {\bibfnamefont{H. A.}\ \bibnamefont{Mook}}, \bibinfo {author} {\bibfnamefont{Y.-Q.}\ \bibnamefont{Wang}}, \bibinfo {author} {\bibfnamefont{B.}\ \bibnamefont{Lorenz}}, \bibinfo {author} {\bibfnamefont{C. W.}\ \bibnamefont{Chu}},\ }%
  \bibinfo {note} {submitted for publication, Dec. 1, 2010}%
  \bibAnnoteFile{NoStop}{ye:10}%
\bibitem{kenzelmann:06}%
  \BibitemOpen
  \bibfield{author}{%
  \bibinfo {author} {\bibfnamefont{M.}\ \bibnamefont{Kenzelmann}}, \bibinfo
  {author} {\bibfnamefont{A.~B.}~\bibnamefont{Harris}}, \bibinfo {author}
  {\bibfnamefont{A.}\ \bibnamefont{Aharony}}, \bibinfo {author}
  {\bibfnamefont{O.}\ \bibnamefont{Entin-Wohlman}}, \bibinfo {author}
  {\bibfnamefont{T.}\ \bibnamefont{Yildirim}}, \bibinfo {author}
  {\bibfnamefont{Q.}~\bibnamefont{Huang}}, \bibinfo {author}
  {\bibfnamefont{S.}~\bibnamefont{Park}}, \bibinfo {author}
  {\bibfnamefont{G.}~\bibnamefont{Lawes}}, \bibinfo {author}
  {\bibfnamefont{C.}~\bibnamefont{Broholm}}, \bibinfo {author}
  {\bibfnamefont{N.}~\bibnamefont{Rogado}}, \bibinfo {author}
  {\bibfnamefont{R.~J.}~\bibnamefont{Cava}}, \bibinfo {author}
  {\bibfnamefont{K.~H.}~\bibnamefont{Kim}}, \bibinfo {author}
  {\bibfnamefont{G.}~\bibnamefont{Jorge}},\ and\ \bibinfo {author}
  {\bibfnamefont{A.~P.}~\bibnamefont{Ramirez}},\ }%
  \bibfield{journal}{%
  \bibinfo {journal} {Phys. Rev. B}\ }%
  \textbf{\bibinfo {volume} {74}},\ \bibinfo {pages} {014429} (\bibinfo {year}
  {2006})%
  \bibAnnoteFile{NoStop}{meddar:09}%
\bibitem{meddar:09}%
  \BibitemOpen
  \bibfield{author}{%
  \bibinfo {author} {\bibfnamefont{L.}\ \bibnamefont{Meddar}}, \bibinfo
  {author} {\bibfnamefont{M.}~\bibnamefont{Josse}}, \bibinfo {author}
  {\bibfnamefont{P.}\ \bibnamefont{Deniard}}, \bibinfo {author}
  {\bibfnamefont{C.}\ \bibnamefont{La}}, \bibinfo {author}
  {\bibfnamefont{G.}\ \bibnamefont{Andr\'e}}, \bibinfo {author}
  {\bibfnamefont{F.}~\bibnamefont{Damay}}, \bibinfo {author}
  {\bibfnamefont{V.}~\bibnamefont{Petricek}}, \bibinfo {author}
  {\bibfnamefont{S.}~\bibnamefont{Jobic}}, \bibinfo {author}
  {\bibfnamefont{M.-H.}~\bibnamefont{Whangbo}}, \bibinfo {author}
  {\bibfnamefont{M.}~\bibnamefont{Maglione}},\ and\ \bibinfo {author}
  {\bibfnamefont{C.}~\bibnamefont{Payen}},\ }%
  \bibfield{journal}{%
  \bibinfo {journal} {Chem. Mat.}\ }%
  \textbf{\bibinfo {volume} {21}},\ \bibinfo {pages} {5203} (\bibinfo {year}
  {2009})%
  \bibAnnoteFile{NoStop}{meddar:09}%
\end{thebibliography}

\end{document}